\documentclass[twocolumn,showpacs,showkeys,preprintnumbers,amsmath,amssymb,aps]{revtex4}
\usepackage{graphicx}
\usepackage{dcolumn}
\usepackage{bm}

\begin{document}
\title{Quantum-noise randomized data-encryption for WDM fiber-optic networks}
\author{Eric Corndorf}
 \email{corndorf@ece.northwestern.edu}
\author{Chuang Liang}
\author{Gregory S. Kanter}
\author{Prem Kumar}
\author{Horace P. Yuen}
 \affiliation{Center for Photonic Communication and Computing\\
       Department of Electrical and Computer Engineering, Northwestern University\\
       2145 Sheridan Road, Evanston, IL 60208}
\date{\today}
\begin{abstract}
We demonstrate high-rate randomized data-encryption through optical
fibers using the inherent quantum-measurement noise of coherent
states of light.  Specifically, we demonstrate 650Mbps data
encryption through a 10Gbps data-bearing, in-line amplified
200km-long line.  In our protocol, legitimate users (who share a
short secret-key) communicate using an $M$-ry signal set while an
attacker (who does not share the secret key) is forced to contend
with the fundamental and irreducible quantum-measurement noise of
coherent states.  Implementations of our protocol using both
polarization-encoded signal sets as well as polarization-insensitive
phase-keyed signal sets are experimentally and theoretically
evaluated.  Different from the performance criteria for the
cryptographic objective of key generation (quantum key-generation),
one possible set of performance criteria for the cryptographic
objective of data encryption is established and carefully
considered.
\end{abstract}

\maketitle
\section{Introduction}\label{intro}
For more than twenty years, physicists and engineers have
investigated quantum-mechanical phenomena as mechanisms to satisfy
certain cryptographic objectives.  Such objectives include user
authentication, bit commitment, key generation, and recently, data
encryption.  To date, the cryptographic objective most considered in
the literature has been key generation.  In key generation, two
users, who initially share a small amount of secret information,
remotely agree on a sequence of bits that is both larger than their
original shared information and is known only to them.  The newly
generated bits (keys) are then used to publicly communicate secret
messages over classical channels by driving data encrypters like the
information-theoretically perfect one-time pad~\cite{Vernam26} or
more efficient (but less secure) encrypters, such as the Advanced
Encryption Standard, where security is described in terms of
complexity assumptions~\cite{Schneier96,Daemen00}.

Several approaches to key generation using quantum effects have been
proposed and demonstrated.  The most famous of these protocols, the
BB84 protocol~\cite{Bennett84} and the Ekert protocol~\cite{Ekert91}
have enjoyed considerable theoretical consideration as well as
experimental implementation~\cite{Gisin02,Tittel00,Jennewein00}.  A
major technical limitation of the BB84 (Ekert) protocol is that the
achievable key-generation rate (more importantly, the rate-distance
product) is relatively low due to the protocol's requirement for
single-photon (entangled-photon) quantum states. This requirement is
a burden not only in the generation of such states, but also in that
such states are acutely susceptible to loss, are not optically
amplifiable (in general), and are difficult to detect at high rates.
Furthermore, because the received light must be detected at the
single-photon level, integration of the protocol implementations
into today's wavelength-division-multiplexed (WDM) fiber-optic
infrastructure is problematic because cross-channel isolation is
typically no better than 30dB.

Recently, we have demonstrated a new quantum cryptographic scheme,
based on Yuen's KCQ approach~\cite{Yuen04}, in which the inherent
quantum noise of coherent states of light is used to perform the
cryptographic service of data
encryption~\cite{Barbosa03,Corndorf03}.  Unlike single-photon
states, coherent states (of moderate average-energy level) are
easily generated, easily detected, and are optically amplifiable,
networkable, and loss tolerant.  Note that key generation and data
encryption are two \emph{different} cryptographic objectives with
\emph{different} sets of criteria by which to judge performance---a
direct comparison between the two is not appropriate.

In our scheme, legitimate users extend a short, shared secret-key by
using a publicity known deterministic function.  The transmitter
uses the extended key to select a signal set for each transmitted
bit such that the legitimate receiver, using the same extended key,
is able to execute a simple binary-decision measurement on each bit.
An eavesdropper, on the other hand, who does not possess the secret
key, is subject to an irreducible quantum uncertainty in each
measurement, even with the use of ideal detectors.  This uncertainty
results in randomization of the eavesdropper's observations, thereby
realizing a true randomized cipher \cite{Massey88} which effectively
limits the eavesdropper's ability to decipher the transmitted
message.  This randomization is ``free" in that it does not require
any additional action on the part of the transmitter in contrast to
other randomized ciphers \cite{Maurer92, Rivest83}, where active
randomization of the signal-set is required by the transmitter.  Our
scheme, running at data-encryption rates up to 650Mbps, uses
off-the-shelf components and is compatible with today's optical
telecommunications infrastructure.  This paper is organized as
follows: in section~\ref{protocol} we outline our quantum-noise
protected data-encryption protocol (call the $\alpha\eta$ protocol),
in section~\ref{sec} we address issues of security and performance,
and in section~\ref{exp} we summarize our experimental results.

\section{Data encryption protocol}\label{protocol}
\begin{figure}
\centering \scalebox{.75}{\includegraphics{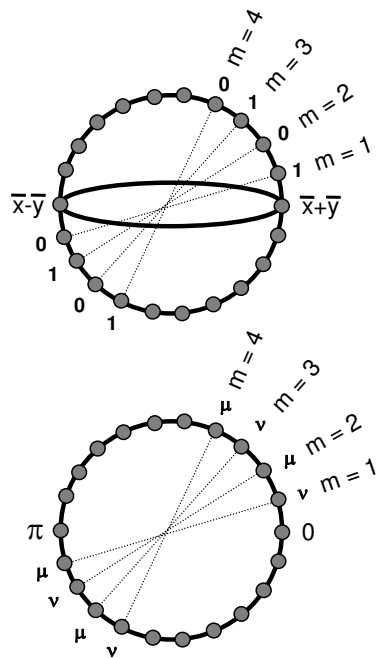}}
\caption{Top: $M$ pairs of orthogonal polarization states uniformly
span a great circle of the Poincar\'{e} sphere; Bottom: $M$ pairs of
antipodal phase states uniformly span a phase circle.}
\label{circles}
\end{figure}

We have implemented two versions of our quantum-noise protected
data-encryption protocol using different signal sets---one using
polarization states~\cite{Corndorf03} (polarization-mode scheme) and
the other using phase states~\cite{Corndorf04-2,Corndorf04}
(time-mode scheme).  In both implementations, the fundamental and
irreducible measurement uncertainty of coherent states is the key
element giving security.  In the polarization-mode scheme, the
two-mode coherent states employed are
\begin{eqnarray}
|\Psi_{m}^{(a)}\rangle&=&|\alpha\rangle_{x}\otimes
|\alpha \,e^{i\theta_{m}}\rangle_{y} \label{eq1}, \\
|\Psi_{m}^{(b)}\rangle&=&|\alpha\rangle_{x}\otimes|
\alpha\,e^{i(\theta_{m}+\pi)}\rangle_{y}\label{eq2},
\end{eqnarray} where $|\alpha\rangle$ is a coherent state, $\theta_{m}=\pi m/M$,
$m \in \{0,1,2,...,(M-1)\}$, $M$ is odd, and the subscripts $x$ and
$y$ indicate the two orthogonal polarization mode-functions.  Viewed
on the Poincar\'{e} sphere, these $2M$ polarization states form $M$
bases that uniformly span a great circle as shown in
Fig.~\ref{circles}(top).  In the time-mode scheme, the single-mode
coherent states employed are
\begin{eqnarray}
|\Psi_{m}^{(a)}\rangle&=&|\alpha e^{i\theta_m}\rangle\label{eq3},\\
|\Psi_{m}^{(b)}\rangle&=&|\alpha e^{i(\theta_m+\pi)}\rangle,\label{eq4}
\end{eqnarray} where again $\theta_{m}=\pi m/M$,
$m \in \{0,1,2,...,(M-1)\}$, and $M$ is odd. These $2M$ states form
$M$ antipodal-phase pairs (bases) that uniformly span the phase
circle, as shown in Fig.~\ref{circles}(bottom).

In both schemes, the transmitter (Alice) extends an $s$-bit secret
key, $\textbf{K}$, to a $(2^{s}-1)$-bit pseudo-random extended-key,
$\textbf{K}'$, using a publicly known $s$-bit linear feedback
shift-register~\cite{Schneier96} (LSFR) of maximal length.  The
extended-key is grouped into continuous disjointed $r$-bit blocks
and then passed through an invertible $r$-bit--to--$r$-bit
deterministic mapping function, referred to as a ``mapper,"
resulting in running-keys, $\textbf{R}$, where $r={\mathrm
{Int}}[\log_{2}M]$ and $s \gg r$. The mapper, which is publicly
known, helps to distribute an attacker's measurement uncertainty
throughout each running-key.  Without the use of a mapper, an
attacker's measurement uncertainty would, the majority of the time,
obscure just a the least-significant bits of each $r$-bit
running-key thereby leaving most of the $r$ bits clearly
identifiable.  Also, note that an LFSR is just one of many of
functions that the users can use to extend $\mathbf{K}$ into
$\mathbf{K'}$. The reason LFSRs are used in these experiments is
because they are mathematically simple to describe which could be
useful when quantifying the precise level of security provided by
$\alpha\eta$.

Depending on the data bit and an instantiation of the running-key
$R$, one of the states in Eqs.~(\ref{eq1}) [(\ref{eq3})] or
(\ref{eq2})[(\ref{eq4})] is transmitted where $m$ is the decimal
representation of $R$.  Specifically, for the polarization-mode
scheme, if $m$ is even then
$(0,1)\rightarrow(|\Psi_{m}^{(a)}\rangle,|\Psi_{m}^{(b)}\rangle)$
and if $m$ is odd then
$(0,1)\rightarrow(|\Psi_{m}^{(b)}\rangle,|\Psi_{m}^{(a)}\rangle)$.
This results in the logical bit mapping of the polarization states
on the Poincar\'{e} sphere to be interleaved $0,1,0,1,...,$ as shown
in Fig.~\ref{circles}(top).  The time-mode scheme is similarly
organized but slightly more complicated in that the data bits are
defined differentially (differential-phase-shift keying, DPSK).
Specifically, if $m$ is even, then the DPSK mapping is
$(0,\pi)\rightarrow(|\Psi_{m}^{(a)}\rangle,|\Psi_{m}^{(b)}\rangle)$,
and
$(0,\pi)\rightarrow(|\Psi_{m}^{(b)}\rangle,|\Psi_{m}^{(a)}\rangle)$
for $m$ odd.  If we relabel the states corresponding to DPSK phases
of ``0" and ``$\pi$" as $\mu$ and $\nu$, respectively, then logical
zero is mapped to $|\Psi_m^{(\mu)}\rangle$
$(|\Psi_m^{(\nu)}\rangle)$ if the previously transmitted state was
from the set $\{|\Psi_m^{(\mu)}\rangle\}$
$(\{|\Psi_m^{(\nu)}\rangle\})$ and logical one is mapped to
$|\Psi_m^{(\mu)}\rangle$ $(|\Psi_m^{(\nu)}\rangle)$ if the
previously transmitted state was from the set
$\{|\Psi_m^{(\nu)}\rangle\}$ $(\{|\Psi_m^{(\mu)}\rangle\})$.  This
results in the mapping of the symbols on the phase circle to be
interleaved $\mu,\nu,\mu,\nu,...,$ as shown in
Fig.~\ref{circles}(bottom).

At the receiving end, the intended receiver (Bob) uses the same
$s$-bit secret key and LFSR/mapper to apply unitary transformations
to his received quantum states according to the running-keys.  These
transformations correspond to polarization-state rotations for the
polarization-mode scheme, and phase shifts for the time-mode
scheme---in either case the transmitted $M$-ry signal set is reduced
to a binary signal-set.  The resulting states under measurement,
depending on the logical bit, are
\begin{eqnarray}
|\Psi^{(a)}\rangle'&=&|\eta\alpha\rangle_{x}\otimes
|\eta\alpha\rangle_{y}, \\
|\Psi^{(b)}\rangle'&=&|\eta\alpha\rangle_{x}\otimes|
-\eta\alpha\rangle_{y},
\end{eqnarray}
for the polarization-mode scheme and
\begin{eqnarray}
|\Psi^{(a)}\rangle'&=&|\eta\alpha\rangle, \\
|\Psi^{(b)}\rangle'&=&|-\eta\alpha\rangle,
\end{eqnarray}
for the time mode scheme, where $\eta$ is the channel
transmissivity. For both schemes the states are then demodulated and
differentially detected.  Specifically, a fixed $\pi/4$ polarization
rotation on the states in the polarization-mode scheme results in
the detected states
\begin{eqnarray}
|\widetilde{\Psi}^{(a)}\rangle&=&|\sqrt{2}\eta\alpha\rangle_{x}\otimes|0\rangle_{y}, \label{p1}\\
|\widetilde{\Psi}^{(b)}\rangle&=&|0\rangle_{x}\otimes|\sqrt{2}\eta\alpha\rangle_{y},\label{p2}
\end{eqnarray}
whereas temporally-asymmetric interferometry in the time-mode
implementation results in the detected states
\begin{eqnarray}
|\widetilde{\Psi}^{(a)}\rangle&=&|\eta\alpha\rangle_{1}\otimes|0\rangle_{2}, \\
|\widetilde{\Psi}^{(b)}\rangle&=&|0\rangle_{1}\otimes|\eta\alpha\rangle_{2}.
\end{eqnarray}

An important feature to note is that Bob does not require high
precision in applying decryption transformations to a transmitted
bit. While the application of a slightly incorrect
polarization/phase transformation results in a larger probability of
error for the bit, it does not categorically render a bit to be in
error. For small perturbations to the polarization/phase rotation,
the majority of the signal energy stays in one of the two detection
modes. The same applies to Bob's detector noise; while an ideal
detector allows for optimized performance, a noisy detector does not
limit Bob's decryption ability beyond an increased probability of
bit error.

\begin{figure*}
\centering \scalebox{.8}{\includegraphics{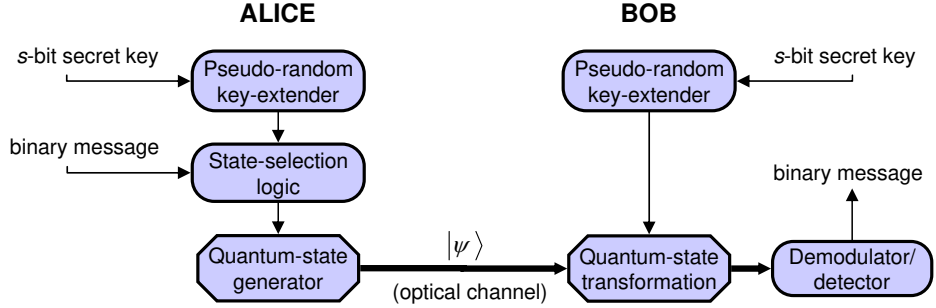}}
\caption{Summary of the quantum-noise protected data encryption
protocol.  In our experiments, the ``pseudo-random key-extender" is
implemented by a maximal-length LFSR and ``$r$-bit--to--$r$-bit
mapping function".} \label{flow}
\end{figure*}

A high-level block diagram of the $\alpha\eta$ protocol is provided
in Fig.~\ref{flow}.  Note that some elements of our protocol that
help to protect the secret key against attack have been
intentionally omitted from this description for the purpose of
clarity.  These omitted elements are mentioned in the following
section and are further described in Ref.~\cite{Yuen04}.
\section{Security}\label{sec}
As stated in the introduction, key generation and data encryption
are different cryptographic objectives and, therefore, have
different sets of criteria by which to evaluate performance.  The
delineation between key generation and data encryption is somewhat
confused by terminology. Because keys procured by a key-generation
protocol are usually assumed to drive deterministic encrypters, the
terms ``quantum key-generation" and ``quantum data-encryption" are
sometimes used interchangeably.  This easily leads to confusion in
that (a) there are potential uses for the generated keys beyond data
encryption, and (b) there are methods of realizing quantum-based
data-encryption without key generation.

In quantum key-generation, a necessary (but not sufficient)
condition that must be satisfied is
\begin{eqnarray}
H(\mathbf{X|Y^E,K})-H(\mathbf{X|Y^B,K})-H(\mathbf{K})>0\label{kg},
\end{eqnarray}
where $\mathbf{X}$ is a classical $n$-bit random vector describing
the transmitted bits, $\mathbf{Y^E}$ and $\mathbf{Y^B}$ are $n$-bit
vectors describing the observations of an attacker (Eve) and Bob,
respectively; $\mathbf{K}$ is an $s$-bit, previously shared secret
between Alice and Bob that might become public on completion of the
protocol, and $H(\cdot)$ is the Shannon entropy function.  Note that
while often omitted in descriptions of the BB84 and Ekert protocols,
both schemes require a secret key $\mathbf{K}$ for the purpose of
message authentication.  Also note that the $H(\mathbf{K})$ term in
Eq.~(\ref{kg}) may be omitted if both a) information about
$\mathbf{K}$ is never publicly announced, and b) $\mathbf{K}$
remains secret even when under a general attack (as in some of
Yuen's KCQ key-generation protocols).

The mathematical definition of $H(\mathbf{\mathbf{X}|\mathbf{Y}})$,
to be read as ``the uncertainty of $\mathbf{X}$ given $\mathbf{Y}$,"
is given by
\begin{eqnarray}
H(\mathbf{X|Y})&\equiv&-\sum_{\mathbf{x,y}}p(\mathbf{X=x},\mathbf{Y=y})\nonumber\\
& &\times\log p(\mathbf{X=x}|\mathbf{Y=y}),\label{H_def1}
\end{eqnarray}
which, with application of Bayes' theorem and the Law of Total
Probability, becomes
\begin{eqnarray}
H(\mathbf{X|Y})&=&-\sum_{\mathbf{x,y}}p(\mathbf{X=x})p(\mathbf{Y=y}|\mathbf{X=x})\nonumber\\
&
&\times\log\left[\frac{p(\mathbf{X=x})p(\mathbf{Y=y}|\mathbf{X=x})}{\sum_{\mathbf{x'}}
p(\mathbf{X=x'})p(\mathbf{Y=y}|\mathbf{X=x'})}
\right].\nonumber\\\label{H_def2}
\end{eqnarray}
The conditional probability distribution $p(\mathbf{Y}|\mathbf{X})$
is completely and uniquely specified by the probability distribution
of the secret key $p(\mathbf{K})$, the probability distribution of
the plaintext message  $p(\mathbf{X})$, and the encryption function
that takes $\mathbf{X}$ to $\mathbf{Y}=E_{\mathbf{K}}(\mathbf{X})$.
While $E_\mathbf{K}(\mathbf{X})$ is usually assumed known to the
attacker via the \emph{Kerckhoff assumption}, it is important to
emphasize that the calculation of $H(\mathbf{X}|\mathbf{Y})$ also
depends on the probability distributions $p(\mathbf{K})$ and
$p(\mathbf{X})$ according to Eve.  This means that Eve's conditional
entropy $H(\mathbf{X}|\mathbf{Y})$ may change if Eve's probability
distribution $p(\mathbf{X})$ changes due to the acquisition of some
side-information (such as the language of the plaintext message).

For the cryptographic objective of data encryption, be it classical
or quantum-noise--protected, some relevant information-theoretic
quantities are:
\begin{eqnarray}
{\mathrm{i}})& &H(\mathbf{X}|\mathbf{Y^B,K}),\\
{\mathrm{ii}})& &H(\mathbf{X}|\mathbf{Y^E}),\label{quant2}\\
{\mathrm{iii}})& &H(\mathbf{K}|\mathbf{Y^E}),\label{quant3}
\end{eqnarray}
where $\mathbf{X}$ is the $n$-bit transmitted message (plaintext),
$\mathbf{Y^{B}}$ and $\mathbf{Y^{E}}$ are Bob's and Eve's $n$-bit
observations of the encrypted plaintext (ciphertext), and
$\mathbf{K}$ is the $s$-bit secret key shared by the legitimate
users.  In words, these quantities describe i) the error rate for
the legitimate users, ii) the secrecy of the data bits when under
attack, and iii) the secrecy of the secret key when under attack.

When launched on either the data bits or the secret key,
cryptographic attacks are normally divided into two categories,
known-plaintext (KPT) attacks and ciphertext-only (CTO) attacks. CTO
attacks correspond to situations where $p(\mathbf{X})$ is uniform,
according to the attacker. In other words, all $2^n$ possible
messages are transmitted with equal probability. A KPT attack
corresponds to all situations where $p(\mathbf{X})$ is nonuniform
including the totally degenerate deterministic case of
chosen-plaintext. Some example KPT attacks include knowledge of the
native language of the message or perhaps some statistical knowledge
of the message content.  While there are clearly varying degrees of
KPT attacks, a CTO attack refers to the specific case of uniform
$p(\mathbf{X})$.

According to information theory~\cite{Cover91,Shannon48},
Eqs.~(\ref{quant2}) and (\ref{quant3}) satisfy the following
inequalities:
\begin{eqnarray}
H(\mathbf{X}|\mathbf{Y^E})\leq H(\mathbf{K}),\label{limit}\\
H(\mathbf{K}|\mathbf{Y^E})\leq H(\mathbf{K}),\label{li2}
\end{eqnarray}
where Eq.~(\ref{limit}) is known as the Shannon
limit~\cite{Shannon49} which is valid when
$H(\mathbf{X}|\mathbf{Y^E},\mathbf{K})=0$ (our data-encryption
protocol operates in a regime where
$H(\mathbf{X}|\mathbf{Y^E},\mathbf{K})\cong0$\footnote{Yuen's KCQ
approach includes schemes for key generation that depend on the fact
that $H(\mathbf{X}|\mathbf{Y_E},\mathbf{K})\neq 0$.  In the regime
in which $\alpha\eta$ operates,
$H(\mathbf{X}|\mathbf{Y_E},\mathbf{K})$ is effectively zero.}). Note
that in $\alpha\eta$, contrary to the case for key generation [cf.\
Eq.~(\ref{kg})], the condition
$H(\mathbf{X}|\mathbf{Y^E},\mathbf{K})>H(\mathbf{X}|\mathbf{Y^B},\mathbf{K})$
need \emph{not} be satisfied.  In fact the opposite is normally true
where an attacker (given the secret key after measurement) has a
lower bit-error rate than the legitimate receiver.  This is the case
when a significant amount of loss and/or additive noise exists
between the two users where it is assumed that the attacker,
performing an adequate quantum measurement, is located near the
transmitter.

The one-time pad encrypter achieves what Shannon called ``perfect
security" which corresponds to $=H(\mathbf{X})$ in the inequality of
Eq.~(\ref{limit}) when $s=n$.  The practical problem with the
one-time pad is that every data bit to be encrypted requires one bit
of key. More ``efficient," albeit less secure, encrypters operate in
the regime where $s\ll n<\infty$, thereby allowing short secret-keys
to encrypt long messages.  A reasonable information-theoretic goal
of such ``imperfect but efficient" encrypters (practical encrypters)
could be to show
\begin{eqnarray}
H(\mathbf{X}|\mathbf{Y^B,K})\rightarrow0,\\
H(\mathbf{X}|\mathbf{Y^E})=\lambda_1\cdot H(\mathbf{K}),\label{pd1}\\
H(\mathbf{K}|\mathbf{Y^E})=\lambda_2\cdot H(\mathbf{K}),\label{pd2}
\end{eqnarray}
where $s\ll n<\infty$ and $\lambda_{1,2}\rightarrow 1$.  It is
extremely important to emphasize that even if
$\lambda_1,\lambda_2\rightarrow0$, there still may exist a large
complexity-based problem of finding the correct $\mathbf{x}$ even
when given $\mathbf{y^E}$, $p(\mathbf{X})$, $p(\mathbf{K})$, and
$E_\mathbf{K}(\mathbf{X})$---it is in this complexity-based limit in
which all of today's commercial deterministic encrypters are
considered.

According to the given information-theoretic criteria, a goal of
practical data encrypters could be to a) drive $\lambda_{1,2}$ as
close to $1$ as possible for a reasonably large $s$ while still
keeping $s\ll n<\infty$; b) attempt to mathematically prove
Eqs.~(\ref{pd1}) and (\ref{pd2}); and c) if conditions (a) and (b)
cannot be met, insure that the computational (search) complexity is
high even when $\lambda_{1,2}\cdot H(\mathbf{K})=0$.  To date, no
practical data encrypter exists for which Eqs.~(\ref{pd1}) and
(\ref{pd2}) can be rigorously proven, for nontrivial $\lambda$, when
under a KPT attack; no significant complexity-based security has
been proven either.

Note that the appropriate information-theoretic criteria by which to
quantify the security of a data encrypter may be different for
different sociological situations.   For example, satisfying the
criteria given in Eqs. (\ref{pd1}) and (\ref{pd2})
($\lambda_{1,2}=1$) may yield security in some situations, but not
in others.  A different set of operationally-meaningful criteria for
the cryptographic objective of data encryption, which does not rely
on Shannon entropy, has been described in Ref.~\cite{Yuen04}.

Towards the goal of satisfying the cryptographic objective of data
encryption, according to any reasonable information-theory--based
criteria, we offer a new approach to data-encryption wherein the
irreducible uncertainty inherent in the quantum measurement of
coherent states of light is used to perform high-speed randomized
encryption that does not sacrifice the data rate. In our protocol
(section~\ref{protocol}), the logical mappings of the symbols are
interleaved (Fig.~\ref{circles}).  While the users (who share a
short secret-key) are able to make simple binary decisions on the
$M$-ry signal set, an attacker (who does not share the secret key)
is left with an irreducible uncertainty in her measurements due to
the quantum fluctuations inherent to coherent states of light.
Specifically, with $M$ and $|\alpha|^2$ in a particular regime,
measurements of neighboring states, on either the Poincar\'{e}
sphere or the phase circle, overlap and obscure one another. To an
attacker, this overlap is equivalent to Alice broadcasting digital
representations of the $M$-ry signal that are then actively
randomized over the signal's closest neighbors in the signal
constellation.  By using coherent states with a relatively weak
amplitude, a similar randomization is achieved through
quantum-measurement noise which requires no active effort on the
part of the transmitter, but still obscures the true identity of the
state called for by the protocol.  Such randomization is realized
through \emph{any} quantum measurement including direct detection,
balanced homodyne/heterodyne detection, and optimal quantum-phase
detection.

\begin{figure*}
\centering
\includegraphics{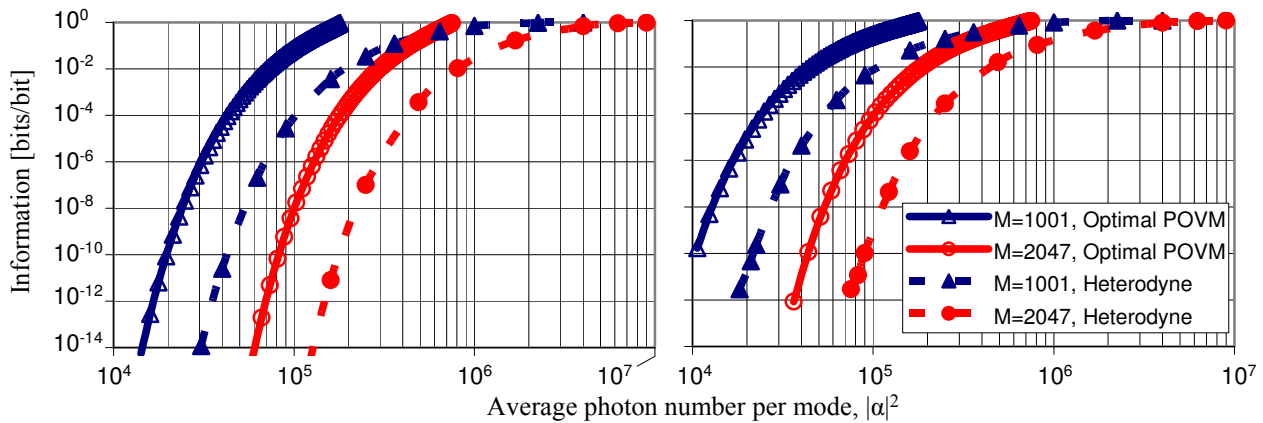}
\caption{Shannon information recovered through individual attacks on
the message when launching either the optimal positive
operator-valued measure or an ideal heterodyne measurement on the
time-mode (left) and polarization-mode (right) implementations.
Plotted as a function of $|\alpha|^2$, for several values of $M$.}
\label{povm_fig}
\end{figure*}

Given some restrictive assumptions, one can even describe the
performance of a quantum-mechanically optimal attack---the best
attack allowed by quantum mechanics.  While the physical structure
of such an  optimal attack may be unknown, quantum mechanics can
establish bounds on the maximum information rate of an attacker. For
individual attacks on the message where classical correlations are
ignored, the quantum-mechanically optimal attack---known as the
optimal positive operator-valued measure---corresponds to optimally
distinguishing all of the states mapped to logical one from those
mapped to logical zero.  Figure~\ref{povm_fig} plots the information
rate of the optimal positive operator-valued measure as a function
of $|\alpha|^2$ and $M$ for the time- and polarization-mode
implementations where information~\cite{Cover91} is defined as
$1+\bar{P}_e \log_2(\bar{P}_e )+(1-\bar{P}_e)\log_2(1-\bar{P}_e)$
for a bit-error rate $\bar{P}_e$.

Figure~\ref{povm_fig} also plots the information rate of the
described attack when performing an ideal heterodyne measurement.
The performance of this measurement is included because it
represents the ``highest performing" receiver structure that an
attacker could practically implement using today's technology.  The
difference between the information rates of the time- and
polarization-mode implementations, for both the optimal positive
operator-valued measure and ideal heterodyne attacks, is due to the
fact that logical bits are defined differentially across two modes
in the time-mode scheme---a bit is correctly determined if and only
if two consecutive state measurements are both correct or both
incorrect. It is important to remember that both the optimal
positive operator-valued measure and ideal heterodyne analyses are
for a very limited attack where Eve does not use her information on
the correlations between the running-keys to determine the plaintext
or secret key---a real attacker would presumably use all information
at her disposal.

While the inability to distinguish neighboring states plays a role
in protecting the secret key against attacks, additional mechanisms
are required to improve the secrecy of the secret key.  By
introducing deliberate state-randomization at the transmitter,
perfect security against CTO attacks on the secret key
[$H(\mathbf{K}|\mathbf{Y^E})=H(\mathbf{K})$, uniform
$p(\mathbf{X})$] can be assured as well as strongly-ideal security
against CTO attacks on the message
[$H(\mathbf{X}|\mathbf{Y^E})=H(\mathbf{K})$, uniform
$p(\mathbf{X})$].  More information on deliberate
state-randomization is available in Ref.~\cite{Yuen04}.  Note that
the mapper and deliberate state-randomization have not yet been
implemented in our published experimental realizations.

Physical ``trojan horse" attacks can also be launched on the message
and the secret key by attempting to \emph{probe} Alice's transmitter
settings.  In such an attack, an eavesdropper would send strong
light into Alice's transmitter and measure the state of her
reflected light.  Attacks of this type can be passively thwarted by
using an optical isolator at the output of Alice's transmitter.

Confusion over the cryptographic service that our protocol
($\alpha\eta$) offers as well as how quantum noise is exploited in
our scheme prompted a criticism~\cite{nishioka04} to
Ref.~\cite{Barbosa03} and some of the authors of
Ref.~\cite{Barbosa03} have replied~\cite{Yuen04-2}.  In
Ref.~\cite{barbosa03-4}, it is claimed that the $\alpha\eta$
data-encryption protocol, operating in a regime where
$H(\mathbf{X}|\mathbf{Y^E},\mathbf{K})<H(\mathbf{X}|\mathbf{Y^B},\mathbf{K})$,
already permits key generation.  We disagree with that conclusion.
\section{Experiments}\label{exp}
Using both the polarization- and time-mode implementations, we
demonstrate high-speed quantum-noise--protected data encryption.
The primary objective of these experiments is to successfully
demonstrate quantum data encryption through a realistic
classical-data bearing WDM fiber line.  A secondary objective is to
show that the quantum-noise encrypted signal does not negatively
impact the performance of the classical data-bearing channels.  The
following two subsections summarize the physical setups as well as
the experimental results for both implementations.
\subsection{Polarization-mode implementation}
\begin{figure*}
\centering
\includegraphics{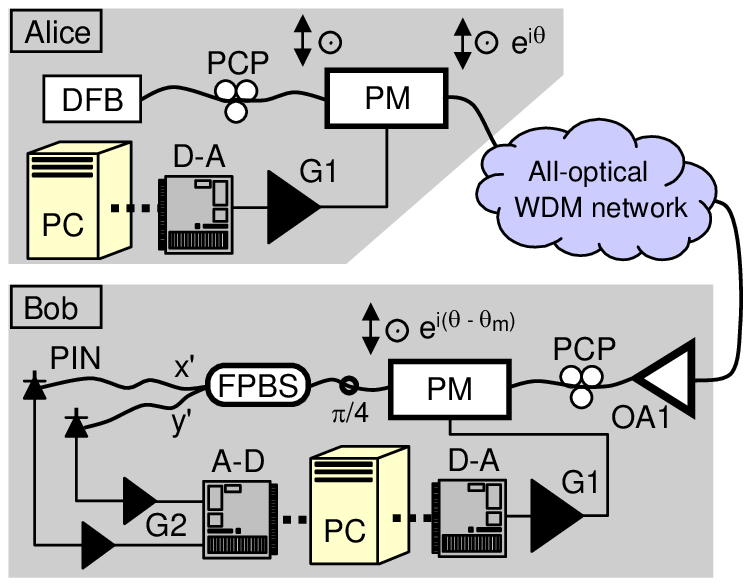}\hskip .5in \includegraphics{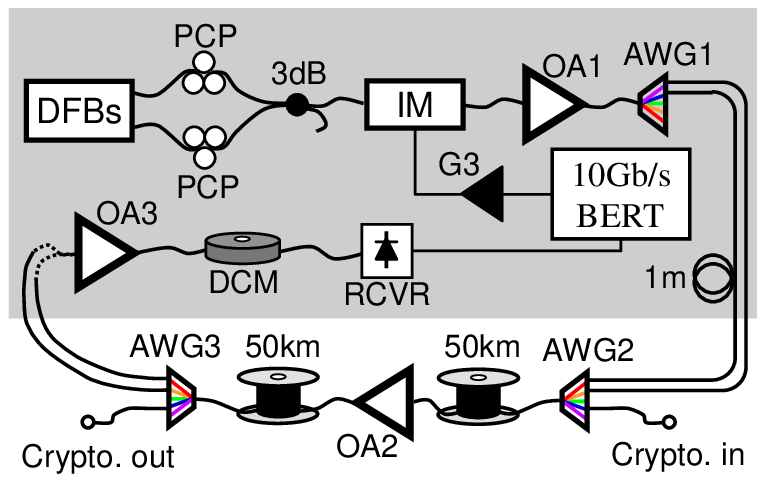}
\caption{Left, transmitter/receiver setup: G1, RF power amplifier;
OA1, low-noise EDFA followed by a Bragg-grating filter; G2, RF
signal amplifier. Right, WDM network setup: OA1, low-noise EDFA; G3,
IM driver; OA2, in-line EDFA followed by an optical isolator; OA3,
EDFA. } \label{exp_pol}
\end{figure*}

A description of the polarization-mode experimental setup naturally
breaks into two parts: the quantum-noise--protected data-encryption
transmitter/receiver pair and the WDM fiber line (which also carries
classical data traffic) over which the encrypted data travels.  We
first describe the transmitter/receiver pair.  As illustrated in
Fig.~\ref{exp_pol}(left), a polarization-control-paddle (PCP) is
adjusted to project the light from a 1550.1nm-wavelength
distributed-feedback (DFB) laser equally into the two polarization
modes of Alice's 10GHz-bandwidth fiber-coupled LiNbO$\rm_{3}$ phase
modulator (PM).  Driven by the amplified output of a 12-bit
digital-to-analog (D-A) board, the modulator introduces a relative
phase ($0$ to $2\pi$ radians) between the two polarization modes.  A
software LFSR, which is implemented on a personal computer (PC),
yields a running-key that, when combined with the data bit,
instructs the generation of one of the two states described in
Eqs.~(\ref{eq1}) and (\ref{eq2}).  Due to electronic bandwidth
limitations of some amplifiers, Manchester coding is applied on top
of the signal set that results in a factor of two reduction of the
data rate (250Mbps) relative to the line rate (500Mbps). Note that
in the time-mode implementation, described in Sec.~\ref{tmode}, such
Manchester coding is not required due to the use of wider bandwidth
amplifiers.

On passing through the 100km-long WDM fiber line [shown in
Fig.~\ref{exp_pol}(right), \emph{Crypto.\ in} and \emph{Crypto.\
out}], the received light is amplified by a home-built
erbium-doped--fiber amplifier (EDFA) with $\simeq30$dB of
small-signal gain and a noise figure very close to the quantum limit
($\rm{NF}\simeq3$dB). Before passing through Bob's PM, the received
light is sent through a second PCP to cancel out the unwanted
polarization rotation that occurs in the 100km-long fiber line.
While these rotations fluctuate with a bandwidth on the order of
kilohertz, the magnitude of the fluctuations drops quickly with
frequency, allowing the use of a manual PCP to track out such
unwanted polarization rotations.  In future implementations Bob's
measurements could be used to drive an automated feedback control on
the PCP.

The relative phase shift (polarization rotation) introduced by Bob's
modulator is determined by the running-key $R$ generated through a
software LFSR in Bob's PC and applied via the amplified output of a
second D-A board.  After this phase shift has been applied, the
relative phase between the two polarization modes is $0$ or $\pi$,
corresponding to a $0$ or $1$ according to the running-key: if $R$
is even then $(0,\pi)\rightarrow(0,1)$ and if $R$ is odd then
$(0,\pi)\rightarrow(1,0)$. With use of a fiber-coupled polarization
beam splitter (FPBS) oriented at $\pi/4$ radians with respect to the
modulator's principal axes, the state under measurement
[Eq.~(\ref{p1}) or (\ref{p2})] is direct-detected by using two
1GHz-bandwidth InGaAs PIN photodiodes operating at room temperature,
one for each of the two polarization modes. The resulting
photocurrents are amplified by a 40dB-gain amplifier, sampled by an
analog-to-digital (A-D) board, and stored for analysis.  The overall
sensitivity of Bob's preamplified receiver is measured to be 660
photons/bit for $10^{-9}$ error probability.

As shown in Fig.~\ref{exp_pol}(right), the 100km-long WDM line
consists of two 40-channel 100GHz-spacing arrayed-waveguide gratings
(AWGs), two 50km spools of single-mode fiber (Corning, SMF-28), and
an in-line EDFA with an output isolator.  Along with the
quantum-noise protected 0.25Gbps encrypted-data channel, two 10Gbps
channels of classical data traffic also propagate through the
described WDM line.  Light from two DFB lasers on the 100GHz ITU
grid (1546.9nm and 1553.3nm) is mixed on a 3dB coupler, where one
output is terminated and the other enters a 10GHz-bandwidth
fiber-coupled Mach-Zender type LiNbO$\rm_{3}$ intensity modulator
(IM).  The IM is driven by an amplified 10Gbps pseudo-random bit
sequence (PRBS) generated by a pattern generator of
$(2^{31}\!\!-\!\!1)$ period.  The PRBS modulated-channels (hereafter
referred to as PRBS channels) then pass through an EDFA to
compensate for losses before entering, and being spectrally
separated by AWG1.  By introducing approximately one meter fiber
length difference between the separated PRBS channels before
combining  them into the 100km-long WDM line with AWG2, the bit
sequences of the two channels are shifted by 50 bits.  This shift
reduces temporal correlations between the two PRBS channels, thereby
more effectively simulating random, real-world data traffic.  The
100km-long WDM line is loss compensated by an in-line EDFA.  The
10dB power loss in the first 50km spool of fiber (0.2dB/km loss) is
compensated by 10dB of saturated gain from the in-line EDFA.  The
overall loss of the line is therefore 15dB, where 10dB come from the
second 50km spool of fiber and the remaining 5dB from the two AWGs
(2.5dB each).

After propagating through the WDM line the channels are separated by
AWG3.  Either of the two PRBS channels is amplified with a 20dB gain
EDFA (OA3) and the group-velocity-dispersion (GVD) is compensated by
a $-1530$ps/nm dispersion-compensation module (DCM).  While the GVD
introduced in the WDM line is approximately 1700ps/nm, the DCM used
is sufficient for our demonstration.  The amplified, GVD-compensated
PRBS channel is detected using an InGaAs PIN-TIA receiver (RCVR) and
analyzed for errors by a 10Gbps bit-error-rate tester (BERT).
Bit-error rates for each PRBS channel are measured separately using
the BERT.

\begin{figure}
\centering
\includegraphics{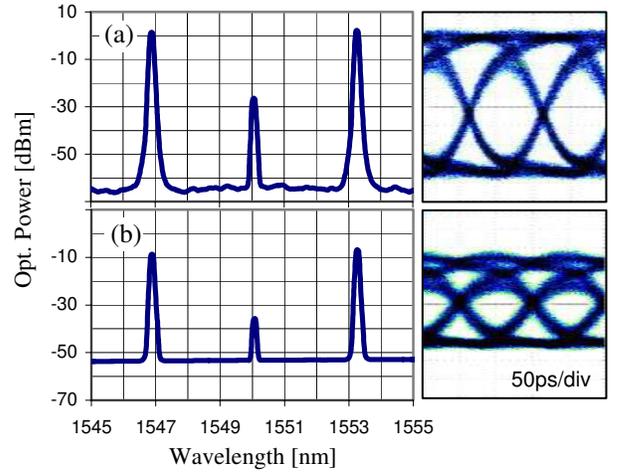}
\caption{(a): Optical spectrum (left) and eye pattern (right) of a
PRBS channel at launch [after AWG2 in Fig.~\ref{exp_pol}(right)].
(b): Optical spectrum (left) and eye pattern (right) of a PRBS
channel at the end of the line [before AWG3 in
Fig.~\ref{exp_pol}(right)].} \label{prbs_pol}
\end{figure}

Figure \ref{prbs_pol}(a)(left) shows the optical spectrum of the
light after AWG2 measured with 0.01nm resolution bandwidth.  The
launch powers in the quantum channel and in each of the PRBS
channels are $-25$dBm and 2dBm, respectively.  An eye pattern of the
1546.9nm PRBS channel at launch is shown in
Fig.~\ref{prbs_pol}(a)(right).  Measuring after AWG2 (i.e., at
launch), neither PRBS channel showed any error in 10 terabits of
pseudo-random data communicated.  Figure \ref{prbs_pol}(b)(left)
shows the optical spectrum (0.01nm resolution bandwidth) of the
light received after the second 50km spool of fiber.  This figure
clearly shows the 10dB loss in signal power of all the channels and
the accompanying 10dB increase in the amplified-spontaneous-emission
dominated noise floor.  An eye pattern of the 1546.9nm PRBS channel,
post dispersion compensation, is shown in
Fig.~\ref{prbs_pol}(b)(right).  While the effect of the residual GVD
is clearly visible in the eye pattern, the bit-error rate for each
of the PRBS channels remains nearly ``error free" at
$5\times10^{-11}$.  Neither the bit-error rates nor the eye patterns
of the PRBS channels change when the quantum channel is turned off.

\begin{figure}
\centering
\includegraphics{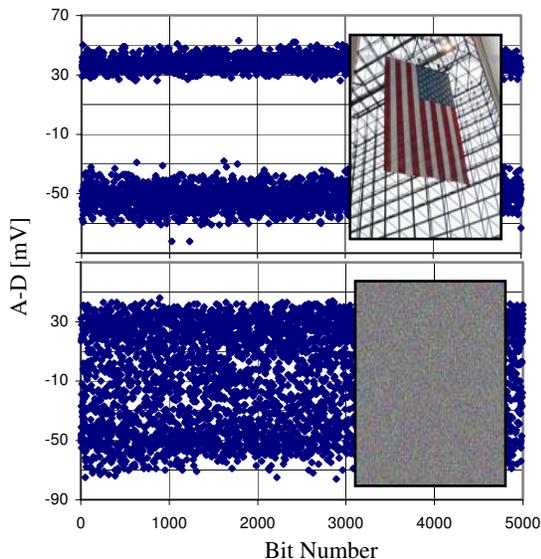}
\caption{5-kbit segments of 9.1-Mbit transmissions through the WDM
link. Insets, the received bit-map images. Top, Bob's detection;
bottom, Eve's detection. } \label{data_pol}
\end{figure}

Figure~\ref{data_pol} shows results of 5000 A-D measurements (one of
the two detector outputs) of a 9.1Mb bitmap file transmitted on the
encrypted channel from Alice to Bob (top) and to Eve (bottom)
through the 100km-long WDM line at 250Mbps data rate. The insets
show the respective decoded images. In this experiment, actions of
Eve are physically simulated by Bob starting with an incorrect
secret-key. Clearly, a real eavesdropper would aim to make better
measurements by placing herself close to Alice and implementing a
more optimized quantum measurement. While Fig.~\ref{data_pol} does
not explicitly demonstrate Eve's inability to distinguish logical
ones from zeros, it does, show that a simple bit decision is
impossible.  In the current setup, the 12-bit D-A conversion allows
Alice to generate and transmit 4094 distinct polarization states
($M=2047$ bases).  The numerical calculation used to plot
Fig.~\ref{povm_fig}(right) then shows that for $-25$dBm launch power
at 250Mbps (500Mbps line rate, $|\alpha|^2\approx20,000$) and
$M=2047$, Eve's maximum obtainable information in an individual
attack on the message is less than $10^{-14}$ bits/bit.
\subsection{Time-mode implementation} \label{tmode}
While technically possible, as demonstrated above, the
polarization-state alignment required at the receiver by the
polarization-mode scheme makes it much less attractive than a
polarization-insensitive version with equivalent performance.  The
time-mode implementation is \emph{totally} polarization-state
insensitive and is therefore much more desirable for performing
quantum-noise--protected data encryption over real-world WDM
networks.

\begin{figure*}
\centering
\includegraphics{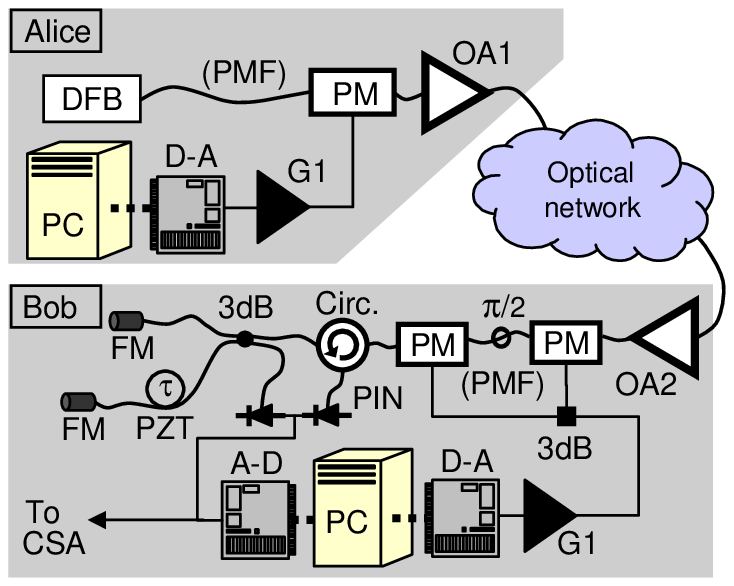}\hskip.5in \includegraphics{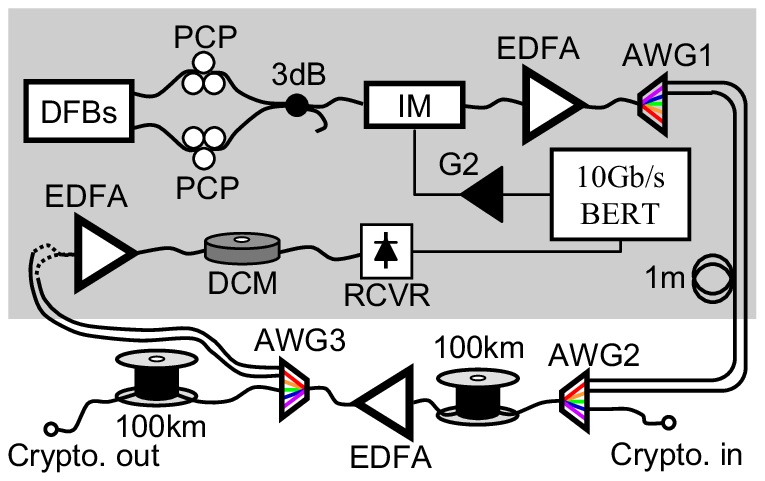}
\caption{Left: Transmitter/receiver setup.  G1, RF power amplifier;
OA2, low-noise EDFA followed by a 25GHz-passband Bragg-grating
filter; PMF, polarization-maintaining fiber; Circ., optical
circulator.  Right: 200km in-line amplified line. IM, 10Gbps
intensity modulator; DCM, dispersion-compensation module; RCVR,
10Gbps InGaAs PIN-TIA optical receiver; G2, 10Gbps modulator driver}
\label{setup_time}
\end{figure*}

As with the polarization-mode implementation, a description of the
time-mode experimental setup naturally breaks into two parts: the
transmitter/receiver pair and the WDM fiber line.  We first describe
the transmitter/receiver pair.  As illustrated in
Fig.~\ref{setup_time}(left), $-25$dBm of power from a
1550.9nm-wavelength DFB laser is projected into Alice's
10GHz-bandwidth fiber-coupled PM.  Driven by the amplified output of
a 12-bit D-A board, the modulator introduces a relative phase ($0$
to $2\pi$ radians) between temporally neighboring symbols.  A 4.4-kb
software LFSR, which is implemented on a PC, yields a running-key
that, when combined with the data bit, instructs the generation of
one of the two states described in Eqs.~(\ref{eq3}) and (\ref{eq4})
at a 650Mbps data rate.  Before leaving the transmitter, the
encrypted signal is amplified with an EDFA (OA1) to a saturated
output power of 2dBm.

On passing through the 200km-long WDM line [shown in
Fig.~\ref{setup_time}(right), \emph{Crypto.\ in} and \emph{Crypto.\
out}), the received light is amplified by another EDFA (OA2) with
$\simeq30$dB of small-signal gain and a noise figure very close to
the quantum limit ($\rm{NF}\simeq3$dB).  The light then passes
through a pair of 10GHz-bandwidth
polarization-maintaining-fiber-coupled PMs oriented orthogonally
with respect to each other so that the $\hat{x}$ $(\hat{y})$
polarization mode of the first modulator projects onto the $\hat{y}$
$(\hat{x})$ mode of the second modulator.  The effect of such
concatenation is to apply an optical phase modulation that is
independent of the polarization state of the incoming light.  The
relative phase shift introduced by Bob's modulator pair is
determined by the running-key $R$ generated through a software LFSR
in Bob's PC and applied via the amplified output of a second D-A
board.  After this phase shift has been applied, the relative phase
between temporally neighboring states is $0$ or $\pi$ (differential
phase-shift keying), differentially corresponding to a $0$ or $1$.

The decrypted signal then passes through a fiber-coupled optical
circulator and into a temporally asymmetric Michelson interferometer
with one bit-period round-trip path-length delay between the two
arms.  Use of Faraday mirrors (FM) in the Michelson interferometer
ensures good polarization-state overlap at the output, yielding high
visibility interference.  The interferometer is path length
stabilized with a PZT and dither-lock circuit.

Light from the two outputs of the interferometer is direct-detected
by using two room temperature 1GHz-bandwidth InGaAs PIN photodiodes
set up in a difference photocurrent configuration.  The resulting
photocurrent is either sampled by an A-D board and stored for
analysis, or put onto a communications signal analyzer (CSA) to
observe eye patterns.

As shown in Fig.~\ref{setup_time}(right), the 200km-long WDM line
consists of two 100GHz-spacing AWGs, two 100km spools of single-mode
fiber (Corning, SMF-28) and an in-line EDFA with an input isolator.
Along with the quantum-noise protected 650Mbps encrypted-data
channel, two 10Gbps channels of classical data traffic also
propagate through the first 100km of the described WDM line.  Light
from two DFB lasers with wavelengths on the 100GHz ITU grid
(1550.1nm and 1551.7nm) is mixed on a 3dB coupler, where one output
is terminated and the other enters a 10GHz-bandwidth fiber-coupled
Mach-Zender type LiNbO$\rm_{3}$ intensity modulator (IM).  The IM is
driven by an amplified 10Gbps PRBS generated by a bit-error-rate
tester (BERT) of $(2^{31}\!\!-\!\!1)$ period.  The PRBS-modulated
channels (hereafter referred to as PRBS channels) then pass through
an EDFA to compensate for losses before entering and being
spectrally separated by AWG1.  Partial decorrelation of the PRBS
channels is achieved by introducing approximately one meter fiber
length difference ($\simeq50$ bits) between the channels before
combining them into the WDM line with AWG2.  On launch (i.e., after
AWG2), the optical power is $-2$dBm/channel for all three channels.

\begin{figure}
\centering
\includegraphics{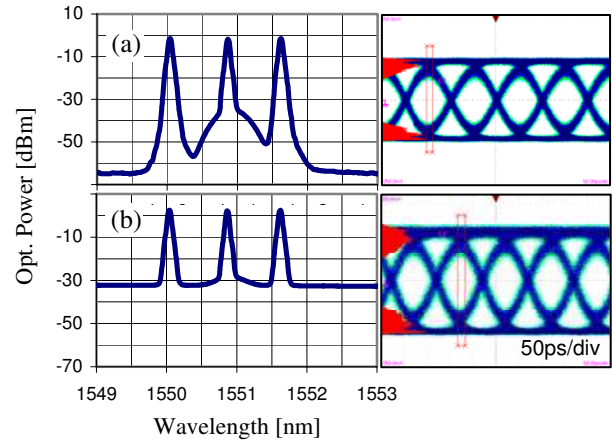}
\caption{(a): Optical spectrum (left) and eye pattern of a PRBS
channel (right) at launch [after AWG2 in
Fig.~\ref{setup_time}(right)].  (b): Optical spectrum (left) and eye
pattern of a PRBS channel (right) after in-line amplification
[before AWG3 in Fig.~\ref{setup_time}(right)].} \label{prbs_time}
\end{figure}

After propagating through the first 100km of fiber (20dB of loss)
and the in-line EDFA (23dB of gain), the channels are separated by
AWG3 (3dB of loss).  Either of the two PRBS channels is amplified
with a 10dB gain EDFA and the GVD is partially compensated by a
$-1530$ps/nm DCM.  The amplified, GVD-compensated PRBS channel is
detected using an InGaAs PIN-TIA receiver (RCVR) and analyzed for
errors by the BERT.  Note that the reason that the PRBS channels do
not propagate through the entire 200km line is because our DCM only
provides enough compensation for 100km of fiber. Figure
\ref{prbs_time}(a)(left) shows the optical spectrum of the light
measured after AWG2 with 0.01nm resolution bandwidth.  The launch
power in the quantum channel and in each of the PRBS channels is
$-1.5$dBm.  An eye pattern of the 1550.1nm PRBS channel at launch is
shown in Fig.~\ref{prbs_time}(a)(right).  Measuring after AWG2
(i.e., at launch), neither PRBS channel showed any errors in 10
terabits of pseudo-random data communicated.  Figure
\ref{prbs_time}(b)(left) shows the optical spectrum (0.01nm
resolution bandwidth) of the light received after the in-line
amplifier (100km of fiber).  An eye pattern of the 1550.1nm PRBS
channel, post dispersion compensation, is shown in
Fig.~\ref{prbs_time}(b)(right).  As in the polarization-mode
implementation, the bit-error rate for each of the PRBS channels
remained nearly ``error free" at $5\times10^{-11}$ despite the
incomplete GVD compensation.  Neither the bit-error rates nor the
eye patterns of the PRBS channels changed when the quantum channel
was turned off.

\begin{figure}
\centering
\includegraphics{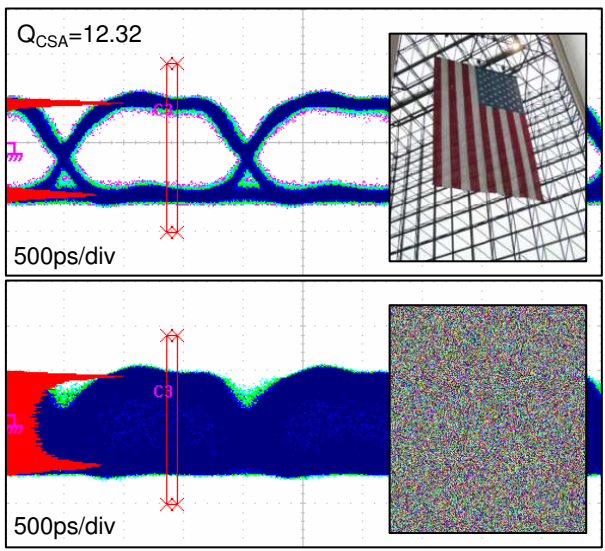}
\caption{Top: Eye pattern and histogram of Bob's decrypted signal
after 200km propagation in the WDM line.  Bottom: Eye pattern and
histogram of Eve's measurements at the transmitter.  Insets,
received 1Mb bitmap file transmissions.} \label{eyes_time}
\end{figure}

Figure~\ref{eyes_time} shows the eye patterns for encrypted 650Mbps
$(2^{15}-1)$-bit-PRBS and 1Mb-bitmap-file transmissions (insets) as
measured by Bob (top) and Eve (bottom).  In these experiments, Bob
is located at the end of the 200km-long line and Eve is located at
the transmitter (Alice).  Eve's actions are physically simulated by
using Bob's hardware, but starting with an incorrect secret-key.
While Fig.~\ref{eyes_time}(bottom) does not explicitly demonstrate
Eve's inability to distinguish neighboring coherent states on the
phase circle, it does, however, show that a simple bit decision is
impossible.  The Q-factor for Bob's eye pattern, as measured on the
CSA, was 12.3.

In all of the time-mode implementation experiments, the coherent
states are transmitted using non-return-to-zero (NRZ) format.  The
return-to-zero-like appearance of Bob's eye pattern is due to
non-zero rise time of the optical phase modulation.  This phenomena
is also observed in traditional NRZ-DPSK systems.  The apparent
banding of Eve's measurements at the top and bottom of the eye
pattern is due to the sinusoidal transfer function of the temporally
asymmetric interferometer used for demodulation.  Despite this
apparent banding, the eavesdropper's probability of error is equal
for every transmitted bit.  If an eavesdropper were to, say, perform
optical heterodyne detection, a uniform distribution of phases would
be observed.

In the current setup, the 12-bit D-A conversion allows Alice to
generate and transmit 4094 distinct phase states ($M=2047$ bases).
Although we simulate an eavesdropper by placing Bob's equipment at
the transmitter, a real eavesdropper would aim to make the best
measurements allowed by quantum mechanics (just as in the
polarization-mode implementation).  The numerical calculation used
to plot Fig.~\ref{povm_fig}(left) shows that for $-25$dBm signal
power at 650Mbps ($\approx40,000$ photons/bit) with $M=2047$, Eve's
maximum obtainable information in an individual attack on the
message would be less than $10^{-15}$ bits/bit.
\section{Discussion and Summary}
In summary, we have developed schemes towards the cryptographic
objective of practical data encryption by using the fundamental and
irreducible quantum-measurement uncertainty of coherent states.
Unlike currently deployed deterministic encrypters whose security
relies solely on unproven computational complexity, we offer a new
quantum-mechanical vehicle to quantifiable information-theoretic
security through high-speed randomized encryption.  Furthermore, we
have clearly specified a set of security criteria for the
cryptographic service of data encryption (which are different from
those for key generation) and considered some optimal quantum
attacks on our scheme.  While we have yet to explicitly determine
the level of information-theoretic security provided by our scheme
under a general attack (which may correspond to finding
$\lambda_1,\lambda_2$), our scheme does provide a physical layer of
quantum-noise randomization that can only enhance the security of a
message already encrypted with a traditional deterministic cipher.

Experimentally, we have implemented and demonstrated two high-speed
versions of the $\alpha\eta$ data-encryption protocol using both
polarization and time modes, and evaluated the schemes' performances
through active WDM lines.  Whereas the polarization-mode experiments
have demonstrated the efficacy of the data-encryption protocol, the
polarization independent time-mode experiments have demonstrated a
technology that is ``drop-in" compatible with the existing optical
telecommunications infrastructure.
\section*{Acknowledgments}
We thank Ranjith Nair and Kahraman G. K\"{o}pr\"{u}l\"{u} for
helpful discussions and analysis.  This work has been supported by
DARPA under grant F30602-01-2-0528.

\end{document}